# LATE-REPLICATING CNVS AS A SOURCE OF NEW GENES

David Juan[1,4], Daniel Rico[1,4,5], Tomas Marques-Bonet[2], Oscar Fernandez-Capetillo[3] and Alfonso Valencia[1]

[1]Structural Biology and BioComputing Programme, Spanish National Cancer Research Centre (CNIO), Madrid, Spain.

[2]Institut Catala de Recerca i Estudis Avancats (ICREA) and Institut de Biologia Evolutiva (UPF/CSIC), Dr Aguader 88, PRBB, Barcelona, Spain.

[3]Genomic Instability Group, Molecular Oncology Programme, CNIO, Madrid, Spain.

[4]These authors contributed equally to this work

[5]Corresponding author: drico@cnio.es

## SUMMARY

Asynchronous replication of the genome has been associated with different rates of point mutation and copy number variation (CNV) in human populations. Here, we explored if the bias in the generation of CNV that is associated to DNA replication timing might have conditioned the birth of new protein-coding genes during evolution. We show that genes that were duplicated during primate evolution are more commonly found among the human genes located in late-replicating CNV regions. We traced the relationship between replication timing and the evolutionary age of duplicated genes. Strikingly, we found that there is a significant enrichment of evolutionary younger duplicates in late replicating regions of the human and mouse genome. Indeed, the presence of duplicates in late replicating regions gradually decreases as the evolutionary time since duplication extends. Our results suggest that the accumulation of recent duplications in late replicating CNV regions is an active process influencing genome evolution.

**INTRODUCTION**

Not all genes in a genome accumulate mutations and evolve at the same rate (Wolfe et al., 1989; Stern and Orgogozo, 2009), a phenomenon for which diverse adaptive and non-adaptive mechanisms have been proposed (Stern and Orgogozo, 2009; Lynch, 2007; Demuth and Hahn, 2009). Recent studies suggest that replication timing (RT) during S-phase may be a non-adaptive factor that contributes to the bias in the accumulation of point mutations (Stamatoyannopoulos et al., 2009; Herrick, 2011; Koren et al., 2012). Indeed DNA replication errors constitute a major source of mutations, which represent the raw material for the evolution of the genome.

The dynamics of replication seems to be largely driven by the configuration of chromatin within the nucleus, whereby more open, physically connected chromosome territories rich in transcriptionally active genes replicate earlier than more tightly packed ones (Hansen et al., 2009; Yaffe et al., 2010; Ryba et al., 2010; De and Michor, 2011). We also know that asynchronous replication of eukaryotic genomes reflects the physical limitations that chromatin compaction exerts on DNA transactions (Ding and MacAlpine, 2011). Late replication of heterochromatic regions of the genome provokes the accumulation of single-stranded DNA (ssDNA), due to the difficulties experienced by DNA polymerase to fill in the gaps. Given that ssDNA is the substrate for recombination reactions that can alter the genome, the accumulation of ssDNA is known as "replication stress" (López-Contreras and Fernandez-Capetillo, 2010). Interestingly, evolutionary divergence and single-nucleotide polymorphisms (SNPs) tend to accumulate in late replicating regions of the human genome, suggesting that during evolution, mutations might have arisen primarily as a consequence of replicative stress (Stamatoyannopoulos et al., 2009). The association between late replication and greater sequence divergence seems to be a general feature of eukaryote genomes and indeed, it has also been reported in the mouse (Pink and Hurst, 2010), yeast (Lang and Murray, 2011) and in flies (Weber et al., 2012).

Whereas point mutations might shape the function of existing genes, the birth of novel genes generally requires mechanisms that generate new genomic regions. Structural changes, such as copy number variants (CNVs), represent one of the main sources of intra- and inter-specific nucleotide differences between individuals (Zhang et al., 2009; Hastings et al., 2009; Mefford and Eichler, 2009). CNVs typically involve intermediate to large regions, providing a substrate for the generation of new genes through gene duplication. Pioneering studies detected pericentromeric and subtelomeric regions as hotspots of segmental duplications and CNVs (Bailey et al., 2001; Mefford and Trask, 2002; Nguyen et al., 2006; Bailey and Eichler, 2006). These regions were clearly enriched in recently expanded gene families, as well as in many repetitive non-coding elements (Horvath et al., 2001). Although other alternative mechanisms have also been proposed (Kaessmann, 2010), copy number variation is thought to be a major source of new genes (Kim et al., 2007; Korbel et al., 2008; Schuster-Böckler et al., 2010).

CNV formation in ancestral species might have led to genomic amplification of regions that contain genes. Later fixation of these regions in the population may occur when a percentage of individuals in a given species harbor a genomic region with an extra gene copy. Although further deletion or pseudogenization might often prevent such genes from becoming fixed (Zhang, 2003; Innan and Kondrashov, 2010), the accumulation of functional genetic changes can eventually lead to the establishment of new genes. An important effect of gene duplication is that evolutionary pressure can be shared between both duplicates due to their initial functional redundancy (Lynch and Force, 2000; Lynch et al., 2001; Zhang, 2003; Innan and Kondrashov, 2010). As a consequence, the duplication event not only creates a new copy of a given gene but also, it may modify the potential mutability of the parental copy, thereby facilitating the exploration of new functional solutions (Ross et al., 2013; Abascal et al., 2013). Interestingly, a significant fraction of the single nucleotide mutations accumulated during genome evolution can be the by-product of the DNA repair low-fidelity mechanisms involved in structural alterations, suggesting a close relationship between point mutations and genomic rearrangements (De and Babu, 2010).

Mechanistically, the models currently used to explain CNV formation involve either non-allelic homologous recombination (NAHR) of (macro or micro) homologous tracks, or non-homologous (NH) repair mechanisms that are at play during replicative stress (e.g., *Fork stalling and Template switching* (FoSTeS) or *Microhomology-mediated break-induced replication* (MMBIR: (Hastings et al., 2009). In humans, CNVs related with NH repair mechanisms are more frequently found in late replicating regions, while NAHR CNVs tend to occur in early replicating regions (Koren et al., 2012). A relationship between late RT and CNV hotspots has also been reported in flies (Cardoso-Moreira et al., 2011). Furthermore, recent data suggest that somatic CNVs in cancer arise as a consequence of replicative stress (Dereli-Öz et al., 2011), and that chromosome structure and RT can be used to predict landscapes of copy number alterations in cancer genomes (De and Michor, 2011). Significantly, chemicals that promote replicative stress increase the rate of *de novo* CNV formation in human immortalized fibroblasts, strong evidence of a mechanistic role for replication stress in the generation of CNVs (Arlt et al., 2009; Arlt et al., 2011).

In this study we aimed to elucidate the possible relevance of the association of CNV regions with later DNA replication times on gene birth and evolution (a scheme representing the different elements analyzed is shown in Figure 1). To address this key question, we followed an approach based on phylostratification, a framework that allows the evolutionary features of protein-coding genes to be identified and studied (Domazet-Loso et al., 2007; Domazet-Lošo and Tautz, 2010; Roux and Robinson-Rechavi, 2011; Chen et al., 2012; Quint et al., 2012). Using this approach we found that RT and copy number variability in protein-coding duplicated genes (PDGs) are radically different depending on their evolutionary age. Our analyses also showed that most human genes duplicated in the Primate lineage are located in late replicating CNV regions. Indeed, this relationship between recent gene duplication and late RT has probably been operating persistently and extensively throughout animal evolution, as we could see that RT parallels gene duplication age in different regions of the human and mouse genome. Our results suggest that molecular features of DNA transactions can influence current genomic structural variations, and that this influence has played a major role in the evolution of the mammalian genome. In particular, these events may facilitate the exploration of new functions through gene birth by duplication, leading to the characteristic distribution of protein function in mammalian genomes.

**RESULTS**

**CNV formation affects evolutionary recent PDGs**

In this work, we studied the potential influence of DNA replication timing on the birth of new genes by duplication in the context of CNVs, recent duplication events that are not fixed but that are spread in populations. CNV regions are likely to be a source of future duplicated genes and evidence is accumulating that suggests their formation is associated to RT (Cardoso-Moreira et al., 2011; De and Michor, 2011; Koren et al., 2012). Therefore, we hypothesized that RT might be a relevant influence on the entire process of CNV generation and gene birth by duplication. Thus, we first examined the relationship between CNVs in human populations and gene duplication during metazoan evolution (Figure 1).

We quantified copy number variation of human protein-coding genes based on CNV maps for 153 humans genomes (Sudmant et al., 2010). Accordingly, we identified genes with CNVs (or CNV-genes) as the 1,092 autosomal protein-coding genes located in regions with either a gain or loss in at least two individuals (see Methods). We explored the association of gene CNV with duplication age (Fig. 1), which was established using a phylostratification protocol (Domazet-Loso et al., 2007). As such, we assigned the evolutionary age of the last duplication in which it was involved to every human protein-coding duplicated gene (PDG) (Roux and Robinson-Rechavi, 2011: see Methods). Duplication events were dated according to 9,432 phylogenetic reconstructions of the 876,985 protein-coding genes from 51 metazoan species and *S. cerevisiae* (Flicek et al., 2011). In this way we were able to distinguish 5,339 protein-coding singleton genes (not duplicated since the appearance of the Metazoa) and 13,985 PDGs within this period of human evolution. Finally, we classified each PDG into 14 *age classes* or phylostrata corresponding to the ancestral species along the timeline of human evolution since the Fungi/Metazoa split (Figure 2A and Table 1, also see Methods). This definition of evolutionary duplication age allows us to analyze the association of different genomic features with the age of the PDGs, helping us to understand the conditions of gene duplication.

We first observed that PDGs as a whole are more often found in human CNV regions than protein-coding singleton genes: 8% and 3%, respectively (P-value = $6.2 \times 10^{-31}$). Having demonstrated a clear association between CNVs and gene duplication, we studied the distribution of CNV genes in different evolutionary duplication ages and we found that recent PDGs are clearly enriched in CNV-genes (Fig 2B; P-value < $10^{-150}$). Indeed, most PDGs duplicated since the primate ancestor were in CNV regions (61%), while most of the genes older than the Eutheria phylostratum (97%) seem to have completely fixed their copy number, which no longer varied in the human population (Fig. 2B).

These results imply that evolutionary recent PDGs are preferentially found in CNV regions, while genes that have not duplicated since the evolution of the first Primates (singletons and older PDGs) are rarely implicated in CNV formation.

**Evolutionary recent PDGs in CNV regions replicate later**

The asynchronous DNA replication that occurs in the genome is related to different patterns of DNA damage, replicative stress and genome rearrangements (Stamatoyannopoulos et al., 2009; Yaffe et al., 2010; De and Michor, 2011; Koren et al., 2012). Most protein-coding gene-rich regions of the genome replicate early. This asymmetric distribution of genes in the genome might somehow reduce the deleterious effects associated to the higher mutation rate in late replicating regions. In this scenario, we decided to investigate if the differences in RT between protein-coding genes were indeed associated to copy number variability.

We calculated the RT of 19,197 human protein-coding genes using the genome-wide RT maps (Ryba et al., 2010) of four different human embryonic stem cell (ESC) lines, which represent the best proxy available for germ-line replication times (Pink and Hurst, 2010). For further analyses we used the *order of replication* of each gene from the genome-wide RT profiles, as a relative measure of the moment of replication of each human protein-coding gene (see Methods). Using these data we found that CNV-genes replicated significantly later than non-CNV genes (Fig. 3A; P-value = $3.4 \times 10^{-15}$). More interestingly, the association of CNV and late replication is distinct for singletons and PDGs. While CNV-PDGs replicate clearly later than non-CNV PDGs (Fig. 3B; P-value = $1.3 \times 10^{-15}$), we did not observe such an association in singleton genes (Fig. 3B; P-value = 0.40).

Based on this observation, we wondered whether this association between CNV PDGs and RT would be even stronger for evolutionary recent genes. This possibility can be explored by differentiating between old and young PDGs (defined as those that duplicated before or after primates evolved). Indeed, we observed a very different behavior for these two age groups (Fig. 3C), whereby recently duplicated PDGs in CNV regions tend to replicate later than young non-CNV PDGs (Fig. 3C, P-value = $3.8 \times 10^{-4}$), a trend that disappears completely for old PDGs (Fig. 3C, P-value = 0.41). These observations were compatible with a prevalent role of CNVs in gene birth through duplication during mammalian evolution. Furthermore, they support the existence of a strong association between recent protein-coding gene duplications and CNV formation in late replicating regions.

**DNA replication timing reflects evolutionary age**

It was evident from our previous analyses (Fig. 3C) that genes duplicated during primate evolution tend to replicate later than older genes (P-value = 3.9 x $10^{-112}$). Thus, we explored the association between gene duplication age and RT in detail, comparing RT in different phylostrata. Strikingly, we observed a clear correlation between RT and gene phylogeny, whereby younger genes gradually became more likely to be replicated later in the S phase (Fig. 4A; rho = 0.21, P-value = 5.1 x $10^{-150}$, Spearman's correlation). This trend is robust, even when we used the RT profiles of human lymphoblasts (Ryba et al., 2010) or fibroblasts (Yaffe et al., 2010) obtained using an alternative methodology (Figure 4–figure supplement 1).

To determine how widespread this correlation was in other mammals, we performed an independent analysis of 14,677 mouse PDGs using mouse ESC RT maps (Hiratani et al., 2009). Following the same phylostratification protocol used for human genes, we classified each mouse PDG according to the 13 age classes associated to the ancestral species in the evolutionary timeline of *Mus musculus* (Table 1 and Figure 4–figure supplement 2). In this way, we again found that the younger mouse PDGs in the mouse genome tend to be late replicating (Fig. 4B; rho = 0.28, P-value = 5.8 x $10^{-278}$, Spearman's correlation). Therefore, the association of gene duplication age and RT appears to be highly significant in Primates and Rodents.

## DNA replication timing reflects evolutionary age at different chromosomal locations

Pericentromeric and subtelomeric regions have previously been described as hotspots of gene duplication (Mefford and Trask, 2002; Bailey and Eichler, 2006) and thus, we evaluated the contribution of these genomic regions to the trends observed in the previous section. We separated the human PDGs into three groups: pericentromeric (1,325 PDGs within 5Mb from the centromere), subtelomeric (2,590 PDGs within 5Mb from the telomere), and interstitial genes (the remaining 15,940 PDGs). Using the same definition, pericentromeric and subtelomeric regions in mouse contain many fewer PDGs (563 and 886, respectively), probably due to the fact that all the autosomal mouse chromosomes are acrocentric, with no protein-coding genes located in the short arms of the chromosome.

We found that PDGs duplicated in the specific human and mouse lineages are significantly enriched at pericentromeric regions of human (P-value = $5.1 \times 10^{-38}$, chi-squared test) and mouse (P-value = $4.4 \times 10^{-5}$) chromosomes. We did not observe a significant enrichment of PDGs duplicated during Primate or Rodent evolution in subtelomeric regions. However, both regions in human are enriched in CNV PDGs, with a 1.44 fold enrichment in subtelomeric regions (P-value = $4.6 \times 10^{-4}$) and 2.35 fold enrichment in pericentromeric regions (P = value $1.6 \times 10^{-19}$). These observations are in agreement to previous estimates (Bailey et al., 2001) and suggest that the contribution of pericentromeric regions to the birth of new duplicates might have been particularly relevant during primates evolution.

We next analyzed the RT of the PDGs in each of the three regions of human chromosomes. The correlation between gene RT and evolutionary age remains statistically significant when human pericentromeric, interstitial and subtelomeric PDGs are analyzed separately (Fig. 5A-B), although it was particularly strong for human pericentromeric PDGs (rho = 0.44, P-value = $1.1 \times 10^{-47}$). We also performed a similar analysis for mouse genes and the association between gene age and RT was also significant for the three chromosomal regions (Fig. 5D-F). In the mouse, the general relationship between RT and gene age was stronger (rho = 0.29, P-value = $5.6 \times 10^{-255}$) in interstitial regions, although it was also significant in the pericentromeric and subtelomeric regions. These observations highlight the prevalence of the association between RT and gene duplication, irrespective of the chromosomal regions where these evolutionary clades concentrate their gene birth events.

In conclusion, the younger the PDG is in evolutionary terms, the later it tends to replicate during S-phase in dividing cells. This surprising temporal parallel can be observed in different mammalian lineages and in different genomic regions. These data reinforce the view of RT as a fundamental element in the organization of the mammalian genome. Remarkably, this relationship can still be detected in the PDGs duplicated at different periods before the mammalian split (P-value = 0.02), suggesting that difficulties associated with late replication (such as replicative stress) might have exerted a strong influence on the evolution of new functions from the earliest stages in the evolution of multicellular organisms.

**DISCUSSION**

We have shown here that protein-coding genes duplicated in evolution (PDGs) are preferentially located in CNV regions. These CNV PDGs are prone to replicate later than non-CNV PDGs, suggesting a link between CNVs, gene duplication and late replication in human cells. We performed a precise phylostratification analysis to determine the ancestral species in which each human PDG was duplicated for the last time. PDGs duplicated after the common *Primate* ancestor were seen to be much more likely to be located in human CNV regions, suggesting that copy number variation in current populations and the fixation of new PDGs are two extremes of a continuous process.

We also observed that *Primate* CNV PDGs replicate even later than *Primate* non-CNV PDGs. This tendency was not observed for older PDGs, which tend to replicate early even if they are located in CNV regions. These results also suggest that copy number formation in gene coding regions is affected distinctly by two mechanisms recently associated to RT. Accordingly, early replicating CNVs are frequently linked to recombination mechanisms such as NAHR, while late replicating CNVs are more frequently associated to non-homology (NH) based mechanisms (Koren et al., 2012) generally associated with replication errors (Hastings et al., 2009). Therefore, singletons and older duplicates that are associated with CNV events would generally be early replicating and involved in recombination events, while CNVs affecting young genes would tend to replicate late as a result of NH mechanisms

Interestingly, we have also shown that RT mirrors the evolutionary age of PDGs in both human and mouse genomes, where younger PDGs tend to replicate later. Indeed, the replication of primate and rodent specific PDGs (protein-coding genes duplicated after the split from their common ancestor) is clearly enriched in the late S-phase. These observations suggest that there is an active process causing newborn duplicated genes to progressively accumulate in the late replicating genomic regions. Although we propose that gene duplication associated to structural variations such as CNVs may be an important factor explaining this trend, retropositions have also been shown to be a source of gene duplicates (Kaessmann, 2010). Given that the trends we observed here are general for all detectable duplicates, future studies will be needed to address the possible differences between duplicates of different origin.

The regular trends observed at distinct evolutionary ages indicate that this process might have been in operation since ancient periods of metazoan evolution. Moreover, this association clearly persists when we analyze pericentromeric, interstitial and subtelomeric regions separately (regions differentially associated to structural variations: Mefford and Trask, 2002; Bailey and Eichler, 2006). These results must be understood in the light of the recently defined "time-invariant principles" of genome evolution (De and Babu, 2010) that refer to aspects of genome evolution that are actually detected at very different time-scales (from cell lifetime to long evolutionary periods). In fact, the parallel between DNA replication and the evolution of gene families by duplication highlights the connection between two processes that occur over extremely different time scales. Eukaryotic DNA replication is completed over approximately 10 hours in dividing human cells, while gene phylogeny represents the accumulated process of gene birth (and loss) over hundreds of millions of years of evolution. In this context, our results indicate that structural and dynamic features of the genome could condition the evolution of its functional organization.

The robustness of the association between duplication age and RT led us to conceptually explore the possible implications of our results in the context of other recent discoveries. It is known that late replicating regions are gene poor in general and particularly deployed of housekeeping/essential genes. In consequence, the insertion of the duplicated material on these regions is very unlikely to be problematic for the new cell. Therefore, the accumulation of new duplicates in these regions could actually facilitate the high rates of gene birth observed in complex species (Prince and Pickett, 2002). In addition, heterochromatin, also defined as *the chromatin that replicates late* (Beisel and Paro, 2011), is a structure clearly associated with late RT, and it can regulate cell type and tissue specific expression. Hence, the chromatin environment in which new genes arise might inherently restrict their expression, thereby reducing their impact on the whole organism while facilitating specific adaptations. This implies that the genomic context where new genes would contribute to the smaller selective pressures found in new genes (Albà and Castresana, 2005; Wolf et al., 2009; Vishnoi et al., 2010).

The preferential birth of new genes in heterochromatic regions provides a platform that might have facilitated, and that would continue to facilitate, rapid evolution in multicellular species (see Fig. 6). In fact, new genes could accumulate mutations faster in late replicating and heterochromatic regions (Stamatoyannopoulos et al., 2009; Pink and Hurst, 2010), since compact chromatin seems to be prone to suffer DNA damage due to replicative stress (Sulli et al., 2012; Alabert and Groth, 2012). At the same time, it is known that DNA damage promotes heterochromatin formation (Jasencakova and Groth, 2010), such that heterochromatin and replicative stress can be considered as both a cause and consequence of each other. Thus, these processes would constitute a feed-forward loop that can contribute to genetic divergence by fueling the birth of new genes and accelerating their evolution. This scenario, where new genes tend to be born in silenced and mutagenic regions could also help understand the accelerated evolution of young genes reported previously (Albà and Castresana, 2005; Wolf et al., 2009; Vishnoi et al., 2010) in terms of a more relaxed selection pressure and of a higher sequence divergence.

In the light of our results and the scheme proposed, the physical limitations on DNA replication and repair that are imposed by the complexity of certain genomic regions might facilitate rapid evolution in eukaryotic cells. However, the potential influence of structural molecular constraints on the evolution of complexity is only just starting to be understood (Prendergast and Semple, 2011; Fernández and Lynch, 2011; Chambers et al., 2013), and the implications of these structural and mechanistic constraints for evolutionary models must still be investigated in depth. Future assessment of the evolutionary relevance of this proposed global scenario will be necessary, and we anticipate that exploring such issues will further advance our understanding of living systems.

**MATERIAL AND METHODS**

**Ensembl and genomic build versions**

We used Ensembl version 61 for all the analyses of the genomic datasets, which corresponds to the human GRCh37.p2 (hg19) and mouse NCBIM37 (mm9) genome builds. We used the Ensembl assembly converter to update the human data in NCBI36 to GRCh37.p2 and the mouse data in NCBIM36 to 37.

**Definition of copy number variable genes**

We used accurate gene copy number variation data from a recent study performed on 159 human genomes (including 15 high coverage genomes: Sudmant et al., 2010). In this study, the authors built genome wide copy number variation (CNV) maps based on a read depth analysis of the corresponding whole-genome shotgun data and they used these maps to estimate the copy number for each individual gene (Sudmant et al., 2010). These authors kindly provided gene copy number estimates for all individuals and 19,315 RefSeq genes. We converted the RefSeq IDs to ENSEMBL IDs using

ENSEMBL-Biomart v61 and we retrieved a total of 17,852 ENSEMBL protein-coding genes with copy number data. The genes smaller than 1 Kb were removed as their copy number estimates are unreliable (Sudmant et al., 2010). We focused on autosomal copy-variable genes, including those genes having 4 or more copies, or less than 2 copies, in at least 2 individuals. Based on these criteria, we obtained a set of 1,092 reliable copy-variable autosomal protein-coding genes.

**Phylostratification of gene duplicates**

We established an analytical pipeline to perform precise phylostratification (Domazet-Loso et al., 2007) in a manner similar to that described recently (Roux and Robinson-Rechavi, 2011). We used the gene family phylogenetic reconstructions of ENSEMBL Compara v61 (Flicek et al., 2011) that are based on genes sequenced from 52 different species. ENSEMBL Compara v61 provides 18,583 annotated gene family trees for 876,985 protein coding genes, and it assigns the speciation or duplication events represented by each internal tree node to the phylogenetic level (or age class) where these events are detected (see Vilella et al., 2009). We used this information in our pipeline to establish the gene duplication age as that of the phylostratum assigned to the last duplication leading to the birth of the extant protein-coding genes. In order to limit the problems associated to reference genomes of species sequenced with low coverage, we only used the age classes defined by species sequenced with relatively high coverage (at least 5X). Singleton genes were defined as those protein-coding genes without a detectable duplication origin in their gene trees.

According to the aforementioned definition of gene duplication age, the age of a protein-coding duplicated gene (PDG) represents that of the ancestral species in which the duplication event that led to the generation of the extant gene was detected.

For this purpose, we only considered duplication events showing a consistency score above 0.3 (Vilella et al., 2009). When this score was exactly 0, we considered that the duplication was an artifact of the phylogenetic reconstruction and we established the gene duplication age in function of the previous node in the tree. Otherwise, we considered the case unclear, such that gene duplication age could not be assigned. Our analysis included the following 14 age classes for human genes: Bilateria, Coelomata, Chordata, Euteleostomi, Tetrapoda, Amniota, Mammalia, Theria, Eutheria (Eutheria + Euarchontoglires), Simiiformes, Catarrhini, Hominidae, HomoPanGorilla and *Homo sapiens* (see Fig. 2 and Table 1). Although there is increasing evidence in support of the still controversial (Huerta-Cepas et al., 2007; Cannarozzi et al., 2007) Euarchontoglires class (Lunter, 2007; Madsen et al., 2001; Murphy et al., 2001), we decided to remove it and to collapse this into the Eutherian level. This is a conservative option due to the inconsistencies described previously between gene trees and species phylogeny at this level (Huerta-Cepas et al., 2007; Cannarozzi et al., 2007). Given that all non-human primate gene builds in ENSEMBL v61 were annotated by projecting human genes from Ensembl v58, we removed all the human genes in ENSEMBL Compara v61 that were not included in Ensembl v58. The mouse PDGs were grouped in the same age classes as the human PDGs from Bilateria to Eutheria, with the addition of the mouse specific lineage classes: Glires, Rodentia, Murinae and *Mus musculus* (see Figure 4–figure supplement 2 and Table 1). Note that only genes duplicated after the Fungal/Metazoan split were classified as PDGs.

**Replication timing in ESCs**

We retrieved the probe log-ratios of the processed and normalized replication times for four human ESCs (BG01, BG02, H7 and H9) from the GEO (Barrett et al., 2011) dataset, GSE20027 (Ryba et al., 2010). These log-ratios were ranked separately for each ESC and each probe log-ratio was substituted by its rank. In order to combine the RT profiles in human ESCs into a unique reference system, we assigned each probe its median rank from the four experiments. For each human protein-coding gene, we assigned the median rank that corresponded to the probe closest to the center of the gene. If the closest probe for a gene was found at a distance further than 10Kb, the gene was no longer considered. All human protein-coding genes were sorted according to these median ranks to estimate the temporal order of replication.

Processed and normalized log-ratios of murine RT correspond to GSE17983 (Hiratani et al., 2009), which contains data for 46C, D3 and TT2 mouse ESCs, were processed in the same manner. The same applies for the RT data from human lymphoblasts (Ryba et al., 2010) and fibroblasts (Yaffe et al., 2010).

**Data processing and statistical analyses**

ENSEMBL databases were accessed using the ENSEMBL Perl API Core and Compara (http://www.ensembl.org/info/docs/api/index.html). The data transformations and file parsing needed to run our gene birth dating pipeline were performed using perl (http://www.perl.org/). All statistical analyses and plots were carried out using R basic functions (http://cran.r-project.org/) and all our code is available upon request.

**AKNOWLEDGEMENTS**

We thank Evan Eichler and Peter Sudmant for sharing the CNV data and for their helpful suggestions. We also thank Manuel Serrano, Federico Abascal and Ramón Díaz-Uriarte for their critical advice; Victor de la Torre, David G. Pisano, Michael L.

Tress, Thomas Glover, James Lupski, Peer Bork and Manel Esteller for helpful discussions; Eduardo A. León for technical help; and the members of the Structural Biology and Biocomputing Programme (CNIO) for interesting comments and support. This work was funded by the BIO2007-66855 project from the Spanish Ministry of Science to A.V. Work in O.F.´s laboratory is supported by grants from the Spanish Ministry of Science (CSD2007-00017 and SAF2011-23753), the European Research Council (ERC-210520), the Association for International Cancer Research (12-0229) and the Howard Hughes Medical Institution (55007417). T.M. is supported by BFU2011-28549 project from the Spanish Ministry of Science and by an ERC Starting Grant (StG_20091118). The authors declare no competing financial interests.

**REFERENCES**

Abascal F, Corpet A, Gurard-Levin ZA, Juan D, Ochsenbein F, Rico D, Valencia A, Almouzni G. 2013. Subfunctionalization via adaptive evolution influenced by genomic context: the case of histone chaperones ASF1a and ASF1b. *Mol Biol Evol* **30:**1853-1866.

Alabert C, Groth A. 2012. Chromatin replication and epigenome maintenance. *Nat Rev Mol Cell Biol* **13:**153-167.

Albà MM, Castresana J. 2005. Inverse relationship between evolutionary rate and age of mammalian genes. *Mol Biol Evol* **22:**598-606.

Arlt MF, Mulle JG, Schaibley VM, Ragland RL, Durkin SG, Warren ST, Glover TW. 2009. Replication stress induces genome-wide copy number changes in human cells that resemble polymorphic and pathogenic variants. *Am J Hum Genet* **84:**339-350.

Arlt MF, Ozdemir AC, Birkeland SR, Wilson TE, Glover TW. 2011. Hydroxyurea induces de novo copy number variants in human cells. *Proc Natl Acad Sci U S A* **108:**17360-17365.

Bailey JA, Yavor AM, Massa HF, Trask BJ, Eichler EE. 2001. Segmental duplications: organization and impact within the current human genome project assembly. *Genome Res* **11:**1005-1017.

Bailey JA, Eichler EE. 2006. Primate segmental duplications: crucibles of evolution, diversity and disease. *Nat Rev Genet* **7:**552-564.

Barrett T, Troup DB, Wilhite SE, Ledoux P, Evangelista C, Kim IF, Tomashevsky M,

Beisel C, Paro R. 2011. Silencing chromatin: comparing modes and mechanisms. *Nat Rev Genet* **12:**123-135.

Cannarozzi G, Schneider A, Gonnet G. 2007. A phylogenomic study of human, dog, and mouse. *PLoS Comput Biol* **3:**e2.

Cardoso-Moreira M, Emerson JJ, Clark AG, Long M. 2011. Drosophila duplication hotspots are associated with late-replicating regions of the genome. *PLoS Genet* **7:**e1002340.

Chambers EV, Rzhetsky A, Bickmore WA, Semple CA. 2013. Divergence of Mammalian Higher Order Chromatin Structure Is Associated with Developmental Loci. *PLoS Comput Biol* **9:**e1003017.

Chen W-H, Trachana K, Lercher MJ, Bork P. 2012. Younger genes are less likely to be essential than older genes, and duplicates are less likely to be essential than singletons of the same age. *Mol Biol Evol* **29:**1703-1706.

De S, Babu MM. 2010. A time-invariant principle of genome evolution. *Proc Natl Acad Sci U S A* **107:**13004-13009.

De S, Michor F. 2011. DNA replication timing and long-range DNA interactions predict mutational landscapes of cancer genomes. *Nat Biotechnol* **29:**1103-8.

Demuth JP, Hahn MW. 2009. The life and death of gene families. *Bioessays* **31:**29-39.

Dereli-Öz A, Versini G, Halazonetis TD. 2011. Studies of genomic copy number changes in human cancers reveal signatures of DNA replication stress. *Mol Oncol* **5:**308-3014.

Ding Q, MacAlpine DM. 2011. Defining the replication program through the chromatin landscape. *Crit Rev Biochem Mol Biol* **46:**165-179.

Domazet-Loso T, Brajković J, Tautz D. 2007. A phylostratigraphy approach to uncover the genomic history of major adaptations in metazoan lineages. *Trends Genet* **23:**533-539.

Domazet-Lošo T, Tautz D. 2010. A phylogenetically based transcriptome age index mirrors ontogenetic divergence patterns. *Nature* **468:**815-818.

number. *Nat Rev Genet* **10:**551-64.

Fernández A, Lynch M. 2011. Non-adaptive origins of interactome complexity. *Nature* **474**:502-505.

Flicek P, Amode MR, Barrell D, Beal K, Brent S, Chen Y, Clapham P, Coates G, Fairley S, Fitzgerald S, et al. 2011. Ensembl 2011. *Nucleic Acids Res* **39:**D800-D806.

Hansen RS, Thomas S, Sandstrom R, Canfield TK, Thurman RE, Weaver M, Dorschner MO, Gartler SM, Stamatoyannopoulos JA. 2009. Sequencing newly replicated DNA reveals widespread plasticity in human replication timing. *Proc Natl Acad Sci U S A* .

Hastings PJ, Lupski JR, Rosenberg SM, Ira G. 2009. Mechanisms of change in gene copy

Herrick J. 2011. Genetic variation and DNA replication timing, or why is there late replicating DNA? *Evolution* **65:**3031-3047.

Hiratani I, Ryba T, Itoh M, Rathjen J, Kulik M, Papp B, Fussner E, Bazett-Jones DP, Plath K, Dalton S, et al. 2009. Genome-wide dynamics of replication timing revealed by in vitro models of mouse embryogenesis. *Genome Res* **20:**155-169.

Horvath JE, Bailey JA, Locke DP, Eichler EE. 2001. Lessons from the human genome: transitions between euchromatin and heterochromatin. *Hum Mol Genet* **10:**2215-2223.

Huerta-Cepas J, Dopazo H, Dopazo J, Gabaldón T. 2007. The human phylome. *Genome Biol* **8:**R109.

Innan H, Kondrashov F. 2010. The evolution of gene duplications: classifying and distinguishing between models. *Nat Rev Genet* **11:**97-108.

Jasencakova Z, Groth A. 2010. Replication stress, a source of epigenetic aberrations in cancer? *Bioessays* **32:**847-855.

Kaessmann H. 2010. Origins, evolution, and phenotypic impact of new genes. *Genome Res* **20:**1313-1326.

Kim PM, Korbel JO, Gerstein MB. 2007. Positive selection at the protein network periphery: evaluation in terms of structural constraints and cellular context. *Proc Natl Acad Sci U S A* **104:**20274-20279.

Korbel JO, Kim PM, Chen X, Urban AE, Weissman S, Snyder M, Gerstein MB. 2008. The current excitement about copy-number variation: how it relates to gene duplications and protein families. *Curr Opin Struct Biol* **18:**366-374.

Koren A, Polak P, Nemesh J, Michaelson JJ, Sebat J, Sunyaev SR, McCarroll SA. 2012. Differential Relationship of DNA Replication Timing to Different Forms of Human Mutation and Variation. *Am J Hum Genet* **91**:1033-1040.

Lang GI, Murray AW. 2011. Mutation rates across budding yeast Chromosome VI are correlated with replication timing. *Genome Biol Evol* **3:**799-811.

López-Contreras AJ, Fernandez-Capetillo O. 2010. The ATR barrier to replication-born DNA damage. *DNA Repair (Amst)* **9:**1249-1255.

Lunter G. 2007. Dog as an outgroup to human and mouse. *PLoS Comput Biol* **3:**e74.

Lynch M, Force A. 2000. The probability of duplicate gene preservation by subfunctionalization. *Genetics* **154:**459-473.

Lynch M. 2007. The frailty of adaptive hypotheses for the origins of organismal complexity. *Proc Natl Acad Sci U S A* **104 Suppl 1:**8597-8604.

Lynch M, O'Hely M, Walsh B, Force A. 2001. The probability of preservation of a newly arisen gene duplicate. *Genetics* **159:**1789-1804.

Madsen O, Scally M, Douady CJ, Kao DJ, DeBry RW, Adkins R, Amrine HM, Stanhope MJ, de Jong WW, Springer MS. 2001. Parallel adaptive radiations in two major clades of placental mammals. *Nature* **409:**610-614.

Marshall KA, Phillippy KH, Sherman PM, et al. 2011. NCBI GEO: archive for functional genomics data sets--10 years on. *Nucleic Acids Res* **39:**D1005-D1010.

Mefford HC, Eichler EE. 2009. Duplication hotspots, rare genomic disorders, and common disease. *Curr Opin Genet Dev* **19:**196-204.

Mefford HC, Trask BJ. 2002. The complex structure and dynamic evolution of human subtelomeres. *Nat Rev Genet* **3:**91-102.

Murphy WJ, Eizirik E, Johnson WE, Zhang YP, Ryder OA, O'Brien SJ. 2001. Molecular phylogenetics and the origins of placental mammals. *Nature* **409:**614-618.

Nguyen DQ, Webber C, Ponting CP. 2006. Bias of selection on human copy-number variants. *PLoS Genet* **2:**e20.

Pink CJ, Hurst LD. 2010. Timing of replication is a determinant of neutral substitution rates but does not explain slow Y chromosome evolution in rodents. *Mol Biol Evol* **27:**1077-1086.

Prendergast JGD, Semple CAM. 2011. Widespread signatures of recent selection linked to nucleosome positioning in the human lineage. *Genome Res* **21:**1777-1787.

Prince VE, Pickett FB. 2002. Splitting pairs: the diverging fates of duplicated genes. *Nat Rev Genet* **3:**827-837.

Quint M, Drost H-G, Gabel A, Ullrich KK, Bönn M, Grosse I. 2012. A transcriptomic hourglass in plant embryogenesis. *Nature* **490:**98-101.

Ross BD, Rosin L, Thomae AW, Hiatt MA, Vermaak D, de la Cruz AFA, Imhof A, Mellone BG, Malik HS. 2013. Stepwise evolution of essential centromere function in a Drosophila neogene. *Science* **340:**1211-1214.

Roux J, Robinson-Rechavi M. 2011. Age-dependent gain of alternative splice forms and biased duplication explain the relation between splicing and duplication. *Genome Res* **21:**357-363.

Ryba T, Hiratani I, Lu J, Itoh M, Kulik M, Zhang J, Schulz TC, Robins AJ, Dalton S, Gilbert DM. 2010. Evolutionarily conserved replication timing profiles predict long-range chromatin interactions and distinguish closely related cell types. *Genome Res* **20:**761-770.

Schuster-Böckler B, Conrad D, Bateman A. 2010. Dosage sensitivity shapes the evolution of copy-number varied regions. *PLoS ONE* **5:**e9474.

Stern DL, Orgogozo V. 2009. Is genetic evolution predictable? *Science* **323:**746-751.

Sudmant PH, Kitzman JO, Antonacci F, Alkan C, Malig M, Tsalenko A, Sampas N, Bruhn L, Shendure J, 1000 Genomes Project, et al. 2010. Diversity of human copy number variation and multicopy genes. *Science* **330:**641-646.

Stamatoyannopoulos JA, Adzhubei I, Thurman RE, Kryukov GV, Mirkin SM, Sunyaev SR. 2009. Human mutation rate associated with DNA replication timing. *Nat Genet* **41:**393-395.

Sulli G, Di Micco R, d'Adda di Fagagna F. 2012. Crosstalk between chromatin state and DNA damage response in cellular senescence and cancer. *Nat Rev Cancer* **12:**709-720.

Vilella AJ, Severin J, Ureta-Vidal A, Heng L, Durbin R, Birney E. 2009. EnsemblCompara GeneTrees: Complete, duplication-aware phylogenetic trees in vertebrates. *Genome Res* **19:**327-335.

Vishnoi A, Kryazhimskiy S, Bazykin GA, Hannenhalli S, Plotkin JB. 2010. Young proteins experience more variable selection pressures than old proteins. *Genome Res* **20:**1574-1581.

Wolf YI, Novichkov PS, Karev GP, Koonin EV, Lipman DJ. 2009. The universal distribution of evolutionary rates of genes and distinct characteristics of eukaryotic genes of different apparent ages. *Proc Natl Acad Sci U S A* **106:**7273-7280.

Weber CC, Pink CJ, Hurst LD. 2012. Late-replicating domains have higher divergence and diversity in Drosophila melanogaster. *Mol Biol Evol* **29:**873-882.

Wolfe KH, Sharp PM, Li WH. 1989. Mutation rates differ among regions of the mammalian genome. *Nature* **337:**283-285.

Yaffe E, Farkash-Amar S, Polten A, Yakhini Z, Tanay A, Simon I. 2010. Comparative analysis of DNA replication timing reveals conserved large-scale chromosomal architecture. *PLoS Genet* **6:**e1001011.

Zhang J. 2003. Evolution by gene duplication: an update. *Trends Ecol Evol* **18:**292 - 298.

Zhang F, Gu W, Hurles ME, Lupski JR. 2009. Copy number variation in human health, disease, and evolution. *Annu Rev Genomics Hum Genet* **10:**451-481.

# FIGURE TITLES AND LEGENDS

**Figure 1. Summary of the analyses performed.**

This figure summarizes the analyses performed in this work, indicating the specific questions addressed and the datasets used. For each human protein-coding duplicated gene (PGD) we determined: (1) its duplication age, (2) whether it is within a CNV region in current human populations, and (3) its replication timing (RT) during S phase. We use this gene-centered information to investigate the involvement of CNVs in gene birth through duplication during human evolution and the possible influence of replication timing in these gene duplication events.

**Figure 2. Phylostratification of human PDGs.**

(A) The age of a duplicated gene represents the ancestral species in which the duplication event that led to the generation of the extant gene was detected. A total of 13,909 gene duplicates were assigned to one of the 14 different evolutionary age groups (or phylostrata). Representative extant species that define the gene age classes are indicated (see Table 1 for the complete list). (B) The proportion of CNV genes in each phylostratum is higher in the genes recently duplicated in evolution (P-value < $10^{-150}$, chi-squared test). A similar result was observed when only CNV gains are considered (Figure 2–figure supplements 1).

**Figure 3. Gene duplications, CNVs and RT.**

(A) The box plots represent the RT of all human protein-coding genes. The RT was obtained from publicly available microarray-based RT maps. A total of 19,197 human genes were ranked from early to late according to their order of replication. Genes located in CNV regions (*CNV genes*) replicate later (P-value = 3.4 x $10^{-15}$, Wilcoxon's test). (B) PDGs in CNV regions replicate later than non-CNV PDGs (P-value = 1.3 x $10^{-15}$), a difference that was not observed for singleton genes (P-value = 0.40). (C) Young PDGs (genes duplicated in the primate phylostrata) are preferentially located in CNV regions that replicate late (P-value = 3,8 x $10^{-4}$, Wilcoxon's test), whereas the difference between CNV and non-CNV PDGs is not significant in older duplicates (P-value = 0.41). Note that PDGs duplicated during Primate evolution tend to replicate later than older genes (P-value = 3.9 x $10^{-112}$). The box width is proportional to the number of genes within each figure panel.

**Figure 4. RT mirrors gene duplication phylogeny.**

(A) RT distribution of human PDGs is correlated with duplication age (rho = 0.21, P-value = 5.1 x $10^{-150}$, Spearman's correlation). (B) RT distribution of mouse PDGs is also correlated with duplication age (rho = 0.28, P-value = 5.8 x $10^{-278}$). The box width is proportional to the number of PDGs within each figure panel, and the specific human and mouse lineage age classes are indicated in bold. See also Figure 4–figure supplements 1-3.

**Figure 5. The association of PDG age and RT is observed in different human and mouse chromosomal regions.**

(A) Human pericentromeric regions (rho = 0.44, P-value = 1.1 x $10^{-47}$, Spearman's rank correlation). (B) Human interstitial regions (rho = 0.18, P-value = 2.7 x $10^{-84}$). (C) Human subtelomeric regions (rho = 0.23, P-value = 5.2 x $10^{-24}$). (D) Mouse pericentromeric regions (rho = 0.17, P-value = 2.0 x $10^{-4}$). (E) Mouse interstitial regions (rho = 0.29, P-value = 5.6 x $10^{-255}$). (F) Mouse subtelomeric regions (rho = 0.32, P-value = 3.6 x $10^{-23}$). Subtelomeric and pericentromeric PDGs were defined as those within 5Mb of the telomere or centromere, respectively. The rest of the PDGs are considered to be in interstitial regions. The box width is proportional to the number of PDGs within each figure panel.

**Figure 6. Proposed model based on our observations and previous knowledge.**

According to our results, a bias in CNV formation (probably associated with replicative stress) leads to the accumulation of CNV-genes in heterochromatin-rich, late replicating regions. This scenario increases the intrinsic probability that new gene copies are located in these regions. In the long term, a recurrence of this situation combined with successive selection events would lead to the progressive accumulation of younger genes in late replicating regions. The location of new genes in heterochromatin would favor the development of cell type-specific patterns of gene expression. This restriction on gene expression will reduce the selection pressure on new genes, resulting in a weaker impact on the whole organism. In this scenario the rapid development of new traits would contribute to the differential evolution of distinct cell types. Obviously, the influence of other unexplored factors would be expected and should not be ruled out.

**TABLES**

**Table 1. List of phylostrata used in the phylogenetic reconstructions.**

| Phylostrata common to Human and Mouse | Human-specific Phylostrata | Mouse-specific Phylostrata | Species |
|---|---|---|---|
| Bilateria | | | *Caenorhabditis elegans* |
| Coelomata | | | *Drosophila melanogaster* |
| Chordata | | | *Ciona intestinalis / Ciona savignyi* |
| Euteleostomi | | | *Tetraodon nigroviridis / Takifugu rubripes / Gasterosteus aculeatus / Oryzias latipes / Danio rerio* |
| Tetrapoda | | | *Xenopus tropicalis* |
| Amniota | | | *Gallus gallus / Meleagris gallopavo / Taeniopygia guttata / Anolis carolinensis* |
| Mammalia | | | *Ornithorhynchus anatinus* |
| Theria | | | *Monodelphis domestica / Macropus eugenii* |
| Eutheria | | | *Vicugna pacos / Tursiops truncatus / Bos taurus / Sus scrofa / Equus caballus / Felis catus / Ailuropoda melanoleuca / Canis familiaris / Myotis lucifugus / Pteropus vampyrus / Erinaceus europaeus / Sorex araneus / Loxodonta africana / Procavia capensis / Echinops telfairi / Dasypus novemcinctus / Choloepus hoffmanni / Mus musculus / Rattus norvegicus / Dipodomys ordii / Cavia porcellus / Spermophilus tridecemlineatus / Oryctolagus cuniculus / Ochotona princeps* |
| | Simiiformes | | *Callithrix jacchus / Tarsius syrichta / Microcebus murinus / Otolemur garnettii / Tupaia belanger* |
| | Catarrhini | | *Macaca mulatta* |
| | Hominidae | | *Pongo pygmaeus* |
| | Homo/Pan/Gorilla | | *Gorilla gorilla / Pan troglodytes* |
| | *Homo sapiens* | | *Homo sapiens* |
| | | Glires | *Ochotona princeps / Oryctolagus cuniculus* |
| | | Rodentia | *Cavia porcellus* |
| | | Murinae | *Rattus norvegicus / Dipodomys ordii / Spermophilus tridecemlineatus* |
| | | *Mus musculus* | *Mus musculus* |

# FIGURE SUPPLEMENT TITLES AND LEGENDS

**Figure 2–figure supplement 1. Proportion of genes affected by CNV-gains in each phylostratum.**

We repeated the analysis shown in Fig. 2B, excluding CNV losses affecting 163 PDGs and considering only CNV-PDGs affected by gains (929). The proportion of genes affected by CNV gains in each phylostratum is higher in the genes recently duplicated in evolution (P-value < $10^{-150}$, chi-squared test).

**Figure 4–figure supplement 1. Replication timing in lymphoblasts and fibroblasts mirrors evolutionary age.**

Analyses equivalent to Figure 4A, using alternative replication timing maps from human lymphoblasts and fibroblasts. (A) RT distribution of human PDGs in lymphoblasts (Ryba et al., 2010) is correlated with duplication age (rho = 0.18, P-value 2.3 x $10^{-103}$, Spearman's correlation). (B) RT distribution of human PDGs is also correlated with duplication age in fibroblasts (profiles obtained using an alternative methodology; rho = 0.17, P-value = 1.6 x $10^{-92}$). The box width is proportional to the number of PDGs, and the specific human and mouse lineage age classes are indicated in bold.

**Figure 4–figure supplement 2. Phylostratification of mouse PDGs.**

The age of a duplicated gene represents the ancestral species in which the duplication event that led to the generation of the extant gene was detected. A total of 14,677 mouse gene duplicates were assigned to one of the 13 different evolutionary age groups (or phylostrata). Representative extant species that define the gene age classes are indicated (see Table 1 for the complete list).

**Figure 4–figure supplement 3. Replication timing in mouse and human ESCs mirrors evolutionary age: alternative representation.**

The distribution of genes in distinct RT fractions of human (A) and mouse (B) phylogenies, using extant species to define the gene age classes as indicated (see Table 1 for the complete list). We grouped all genes into five consecutive temporal clusters that contain a similar number of genes (quintiles S1-S5).

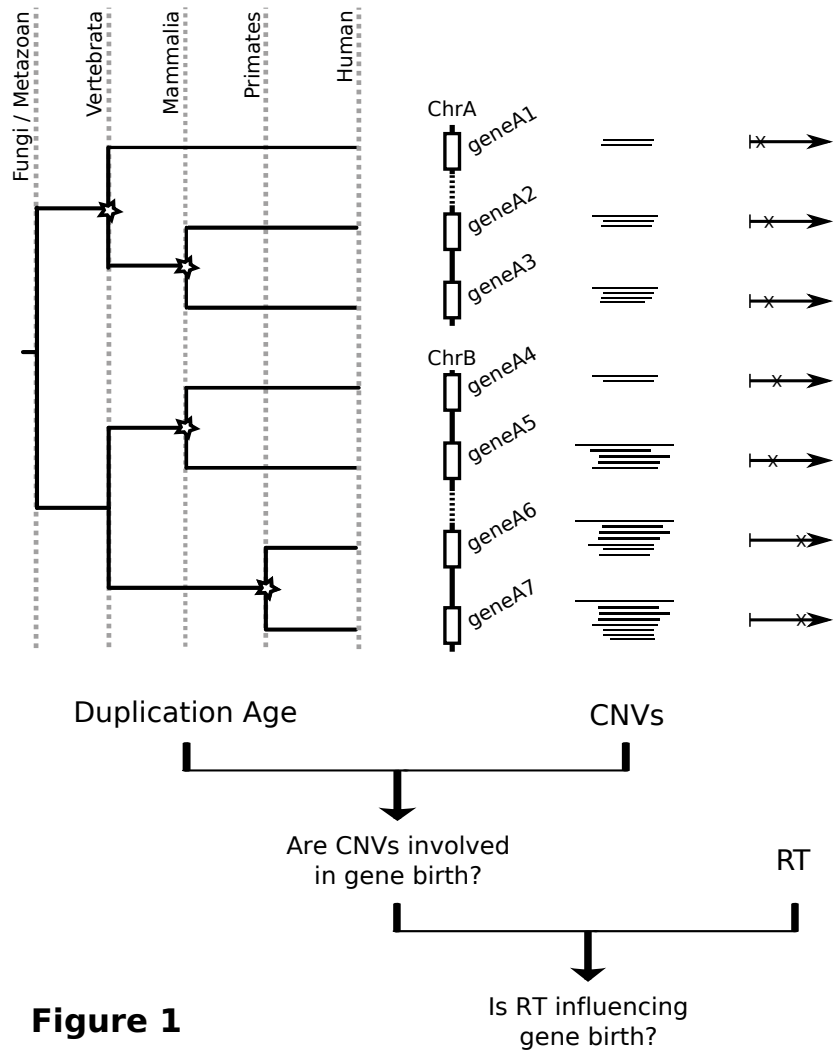

**Figure 1**

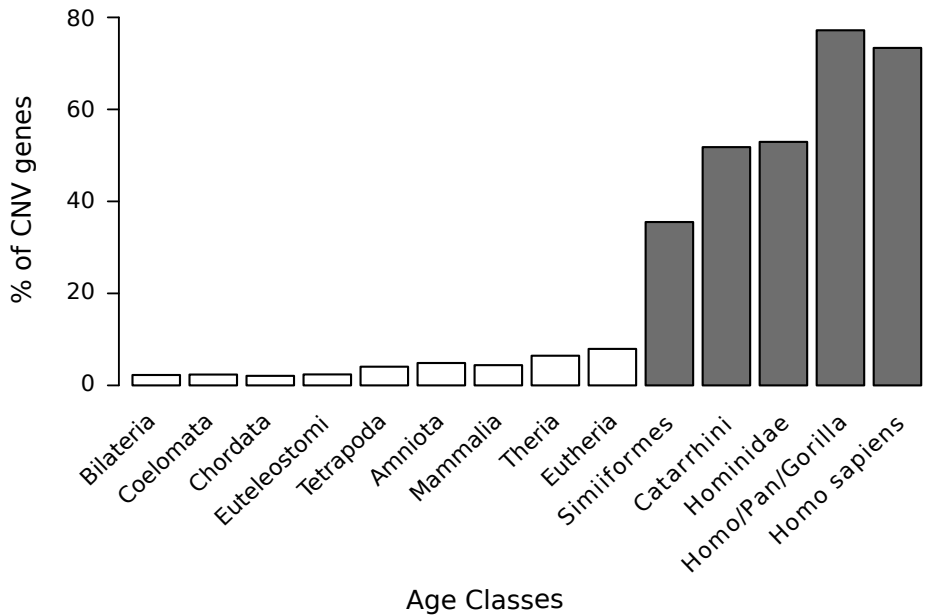

**Figure 2**

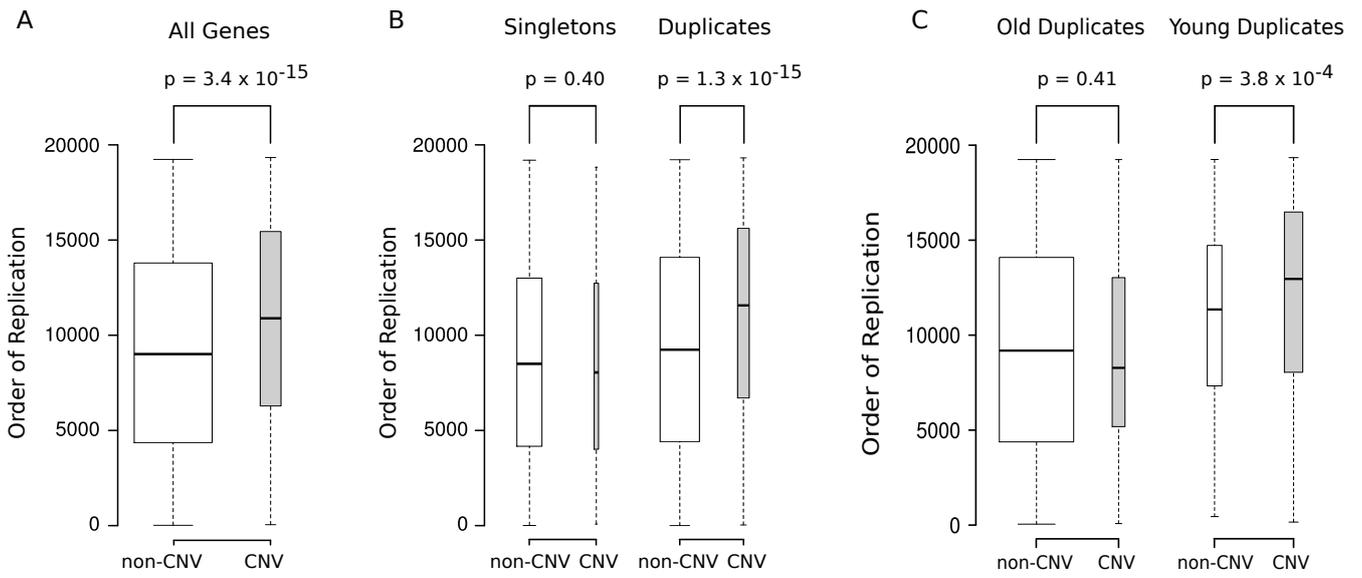

**Figure 3**

A

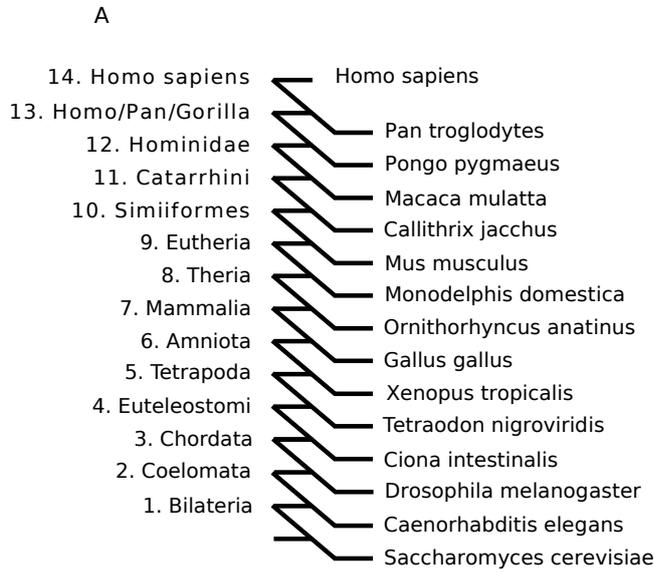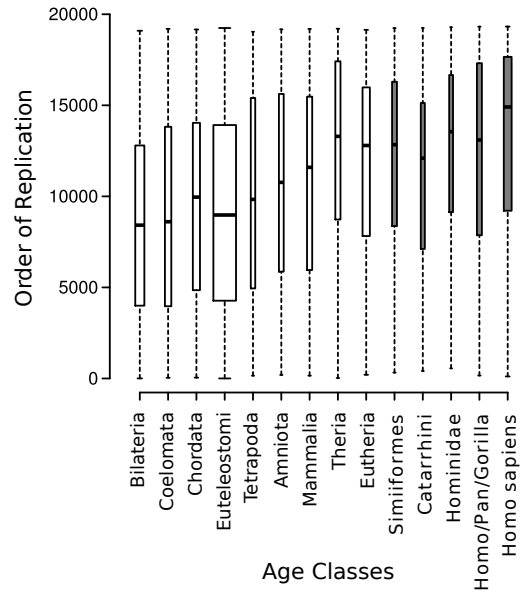

B

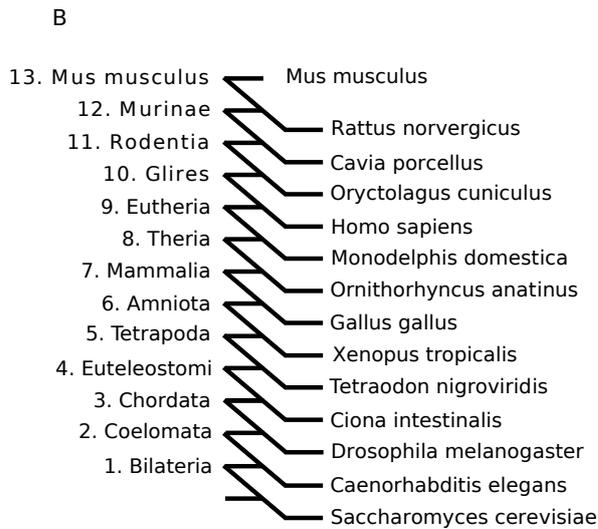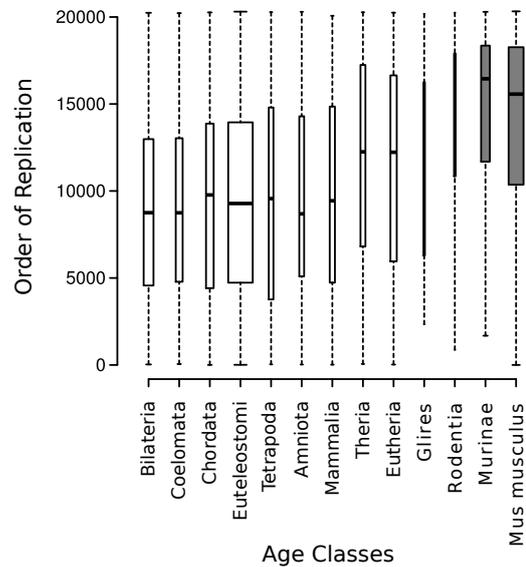

**Figure 4**

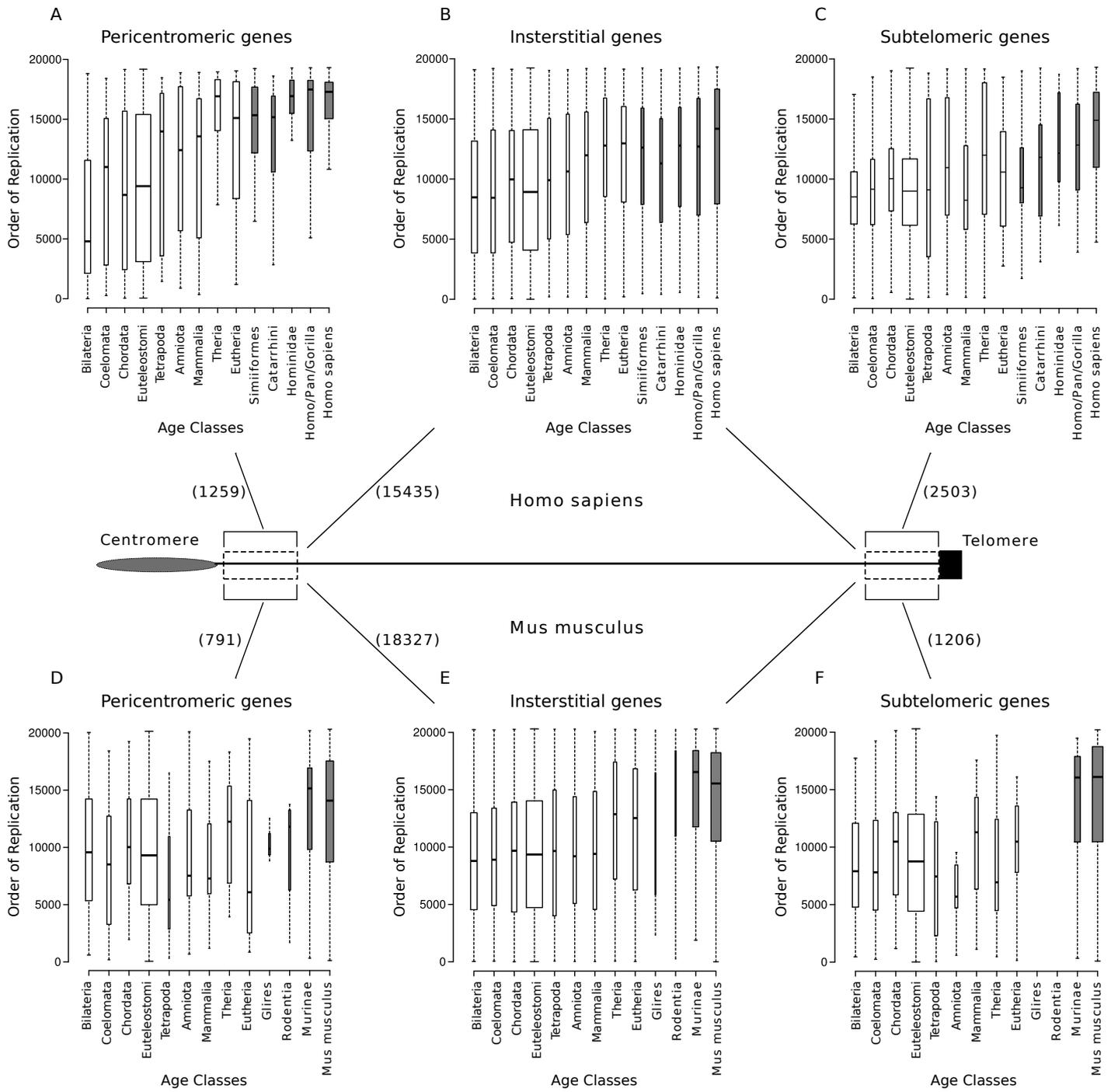

**Figure 5**

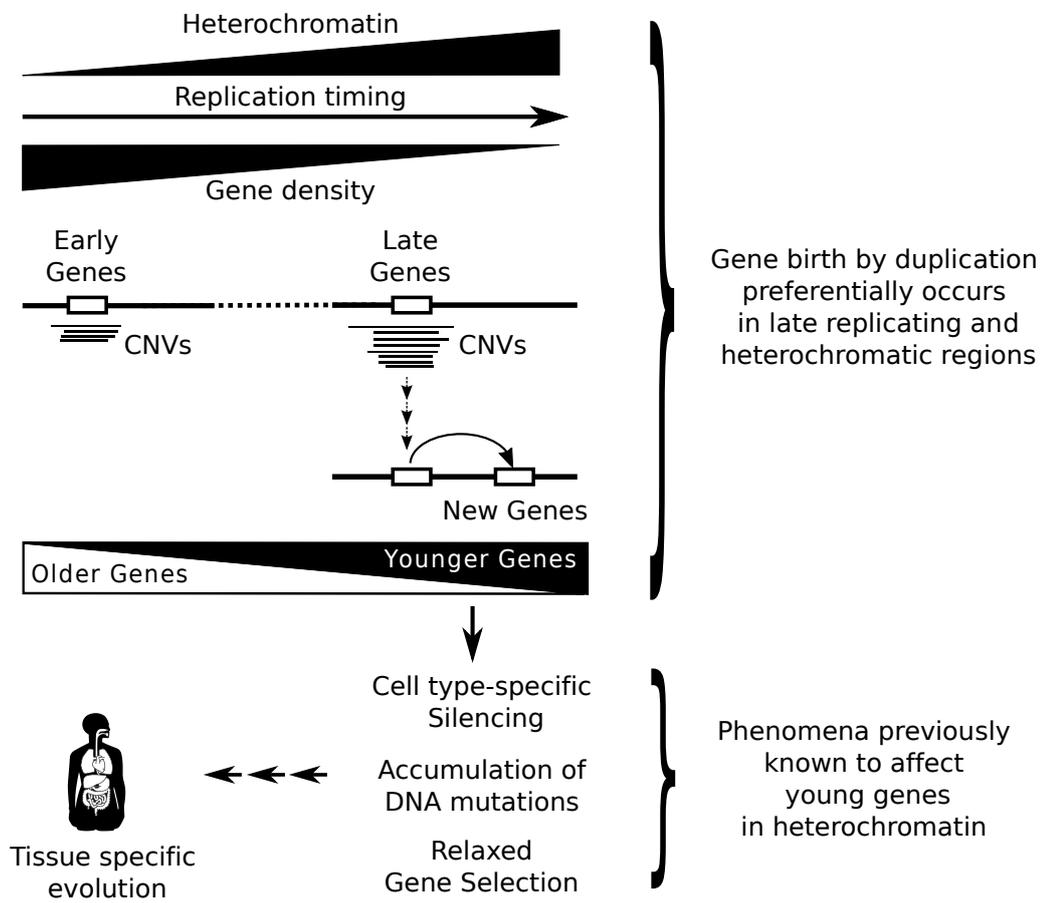

**Figure 6**

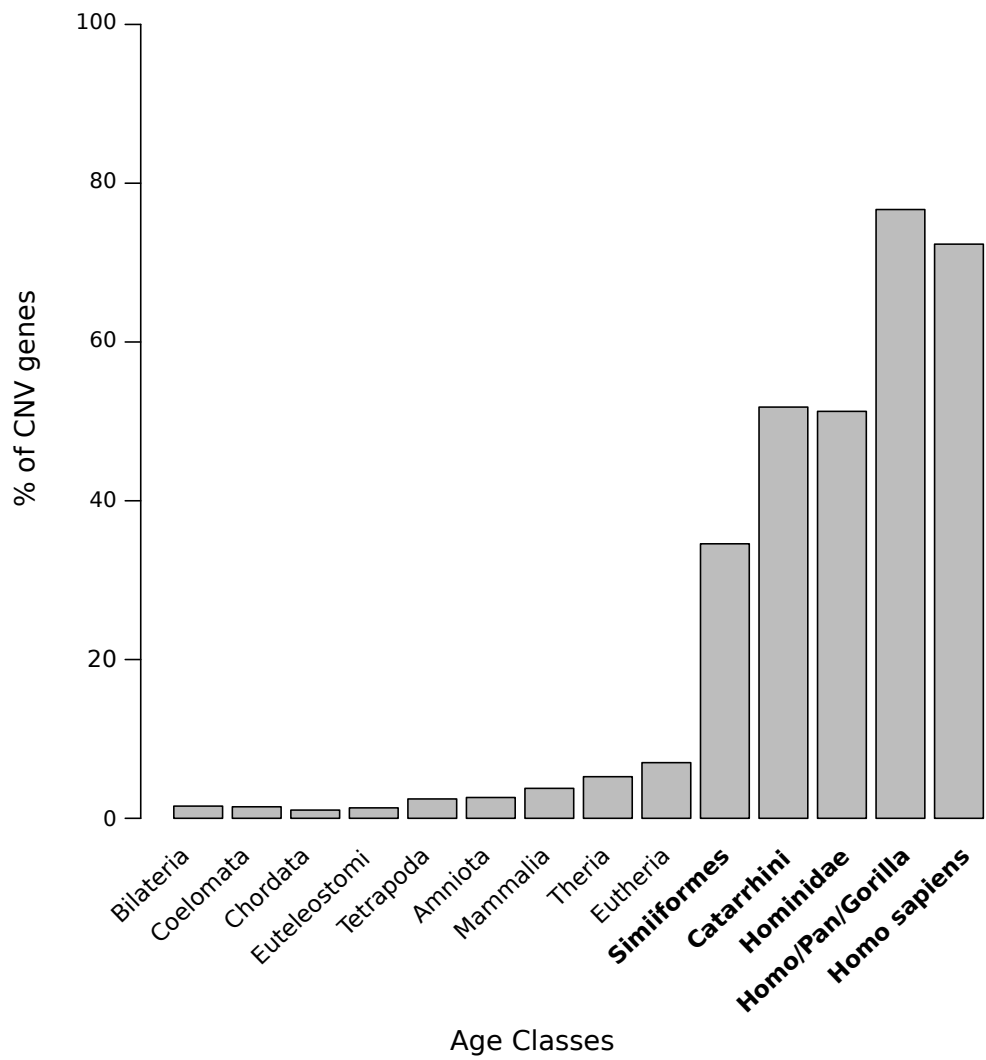

**Figure 2–figure supplement 1**

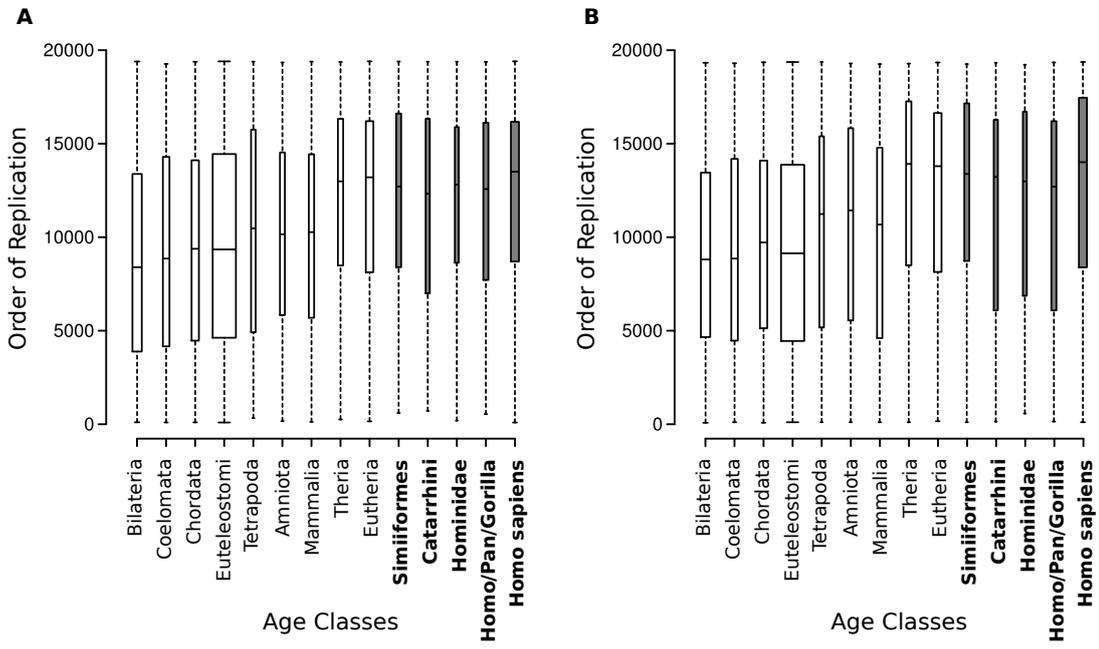

**Figure 4–figure supplement 1**

| Examples of current species | Evolutionary Age classes | Evolutiionary Time Periods (Mya) | Number of duplicates |
|---|---|---|---|
| *Mus musculus* | 13. Mus musculus | 37-0 | 2628 |
| *Rattus norvergicus* | 12. Murinae | 80-37 | 898 |
| *Cavia porcellus* | 11. Rodentia | 81-80 | 30 |
| *Oryctolagus cuniculus* | 10. Glires | 90-81 | 24 |
| *Homo sapiens* | 9. Eutheria | 166-90 | 472 |
| *Monodelphis domestica* | 8. Theria | 184-166 | 287 |
| *Ornithorhyncus anatinus* | 7. Mammalia | 326-184 | 318 |
| *Gallus gallus* | 6. Amniota | 359-326 | 275 |
| *Xenopus tropicalis* | 5. Tetrapoda | 420-359 | 257 |
| *Tetraodon nigroviridis* | 4. Euteleostomi | 550-420 | 7090 |
| *Ciona intestinalis* | 3. Chordata | 570-550 | 690 |
| *Drosophila melanogaster* | 2. Coelomata | 580-570 | 564 |
| *Caenorhabditis elegans* | 1. Bilateria | <~580 | 1144 |
| *Saccharomyces cerevisiae* | | | |

**Figure 4–figure supplement 2**

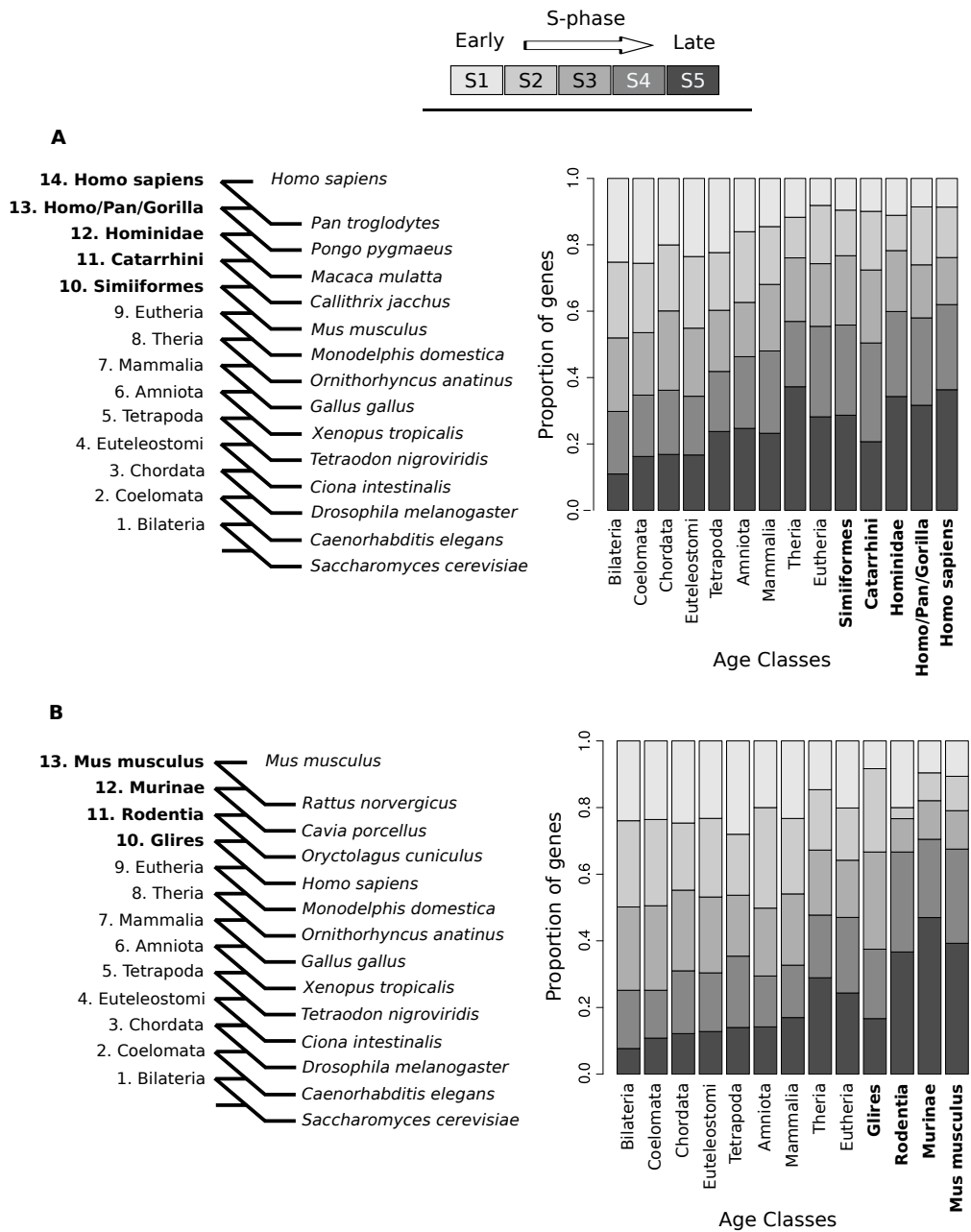

**Figure 4–figure supplement 3**